\documentclass[aip,jap,reprint,graphicx]{revtex4-1}

\usepackage{graphicx}

\begin{document}


\title{Pressure effects on strained FeSe$_{0.5}$Te$_{0.5}$ thin films} 




\author{M. Gooch}
\email{mjgooch@mail.uh.edu}
\author{B. Lorenz}
\affiliation{$TCSUH$ and
Department of Physics, University of Houston, Houston, Texas
77204-5002, USA}
\author{S. X. Huang}
\author{C. L. Chien}
\affiliation{Department of Physics and Astronomy, Johns Hopkins University, Baltimore, Maryland 21218, USA}
\author{C. W. Chu}
\affiliation{$TCSUH$ and
Department of Physics, University of Houston, Houston, Texas
77204-5002, USA and Lawrence Berkeley
National Laboratory, 1 Cyclotron Road, Berkeley, California 94720,
USA}
%


\date{\today}

\begin{abstract}
The pressure effect on the resistivity and superconducting $T_c$ of prestrained thin films of the iron chalcogenide superconductor FeSe$_{0.5}$Te$_{0.5}$ is studied. Films with different anion heights above the Fe layer showing different values of ambient pressure $T_c$'s are compressed up to a pressure of 1.7 GPa. All films exhibit a significant increase of $T_c$ with pressure. The results cannot solely be explained by a pressure-induced decrease of the anion height but other parameters have to be considered to explain the data for all films.
\end{abstract}


\maketitle 

\section{Introduction}
The discovery of superconductivity in iron pnictide compounds has revived superconductivity research with the goal to search for new superconducting compounds with high critical temperatures as well as novel physical phenomena and mechanisms of superconductivity.\cite{kamihara:08,takahashi:08,chen:08b,chen:08,ren:08} With the theoretical proposal of multiband superconductivity with s$_\pm$ pairing symmetry\cite{mazin:08,chubukov:08,cvetkovic:09,wang:09} and the experimental evidence for nodeless two-gap superconductivity in the optimally doped pnictides,\cite{chen:08f,ding:08,matano:08,hashimoto:09,mu:09} it became obvious very soon that a new class of high-temperature superconductors different from the known high-$T_c$ cuprates had been found. Of particular interest is the close proximity of the superconducting state to magnetic order in form of a spin density wave (SDW) state that extends from the underdoped toward to the optimally doped region, in the phase diagram of most iron pnictides. This has raised questions about the role of antiferromagnetic fluctuations in the pairing mechanism of FeAs superconductors.

The new class of superconducting materials is structurally very flexible and many different compounds have been studied (see for example the review of Ref. [15]). The common feature of all is the Fe$_2$As$_2$ layer as the active layer for superconductivity. Unlike the CuO planes in the cuprates, the Fe$_2$As$_2$ layer is not flat but it forms a puckered slab with the planar Fe layer sandwiched between two As sheets. The Fe$_2$As$_2$ layers are separated by different blocks of rare earth and oxygen (LaOFeAs), alkaline earth ions (SrFe$_2$As$_2$), or more complex assemblies of ions.\cite{paglione:10} The Fe-ions are located in the center of a distorted As$_4$ tetrahedron, the shape and dimension of which is considered to be essential for high superconducting temperatures.\cite{lee:08c,zhao:08,mizuguchi:10,okabe:10}

In an attempt to clarify the relationship between the tetrahedral dimensions and high-$T_c$ superconductivity, Huang et al. have recently investigated a series of thin films of the iron chalcogenide FeSe$_{0.5}$Te$_{0.5}$.\cite{huang:10} This compound has the simplest structure and it consists exclusively of Fe$_2$(Se,Te)$_2$ layers isostructural to the Fe$_2$As$_2$ layers, stacked along the $c$-axis of the tetragonal structure. It was demonstrated that high quality $c$-axis oriented films can be deposited on MgO with different substrate temperatures, $T_s$, which determines the biaxial strain in the FeSe$_{0.5}$Te$_{0.5}$ films. The decrease of $T_s$ results in an increase of the $c$-axis and a decrease of the $a$-axis. Most importantly, no chemical substitution or charge doping is necessary and the shape (angle) of the Fe-Se/Te tetrahedron remained approximately constant. The parameter controlling the superconducting $T_c$ was identified as the anion hight above the Fe-layer, $h_{Se/Te}$, that decreases with decreasing substrate temperature. Upon decreasing $h_{Se/Te}$($T_s$), superconductivity arises below a critical value with $T_c$ rapidly increasing toward its maximum that is close to the superconducting $T_c$ of bulk FeSe$_{0.5}$Te$_{0.5}$. The anion height of the films with $T_{c,max}$ is close to the value of $h_{Se/Te}$ for the bulk material. With further decreasing $h_{Se/Te}$ the superconducting $T_c$ falls off quickly and vanishes below another critical value.\cite{huang:10}

The effect of hydrostatic pressure on $T_c$ of iron pnictides and chalcogenides has been studied extensively for bulk compounds.\cite{chu:09} The $T_c$ of bulk FeSe$_{0.5}$Te$_{0.5}$, as defined by the onset of the resistivity drop,  was raised from 14 K at ambient pressure to 19 K at about 3.6 GPa; however, with a significant broadening of the superconducting transition.\cite{horigane:09,huang:09,tsoi:09} With further increasing pressure, $T_c$ did decline quickly with an extrapolated disappearance of superconductivity above 10 GPa. Since the major effect of pressure is the compression of the lattice, it can be anticipated that pressure will also deform the Fe-Se/Te tetrahedra of thin films and have a significant effect on the crucial parameters like the tetrahedral angle or the anion height and consequently modify the superconducting $T_c$. We have therefore studied a series of strained thin films of FeSe$_{0.5}$Te$_{0.5}$ of varying ambient-pressure anion height, deposited at different temperatures on MgO substrates.

\section{Experimental}
Films of thickness 400($\pm$50) nm have been prepared by pulsed laser deposition (PLD) using the same target of composition FeSe$_{0.5}$Te$_{0.5}$, as described in Ref. [20], and four different substrate temperatures have been selected. All films have exclusively c-axis orientation. The composition of the films is close to the target composition since the PLD results in a congruent film deposition and the estimated vapor pressure of FeSe$_{0.5}$Te$_{0.5}$ is smaller than 10$^{-10}$ Torr at 500 $^o$C, well above the chosen substrate temperatures. According to the deposition temperatures the films are labeled as follows: A ($T_s$=240$^o$C), B ($T_s$=290$^o$C), C ($T_s$=340$^o$C), and D ($T_s$=400$^o$C). An additional thin layer of chromium was deposited on top of the films to protect the FeSe$_{0.5}$Te$_{0.5}$ films and to make sure that they are not affected by exposure to air or moisture during the preparation for high-pressure experiments and by the pressure transmitting medium.

Pressure up to 1.7 GPa was generated in a piston-cylinder clamp cell and a 1:1 mixture of Fluorinert FC70:FC77 was used as a liquid pressure medium. The resistivity of the four samples was measured in standard four lead configuration in the plane of the films employing the ac resistance bridge, LR700 (Linear Research). The pressure was measured \emph{in situ} at low temperature by following the superconducting transition of a lead manometer.

\section{Results and discussion}
The ambient pressure resistivity data are shown in Fig. 1. The films A, B, C are superconducting and the fourth sample (D) exhibits a semiconducting temperature dependence of the resistivity above 2 K. The superconducting $T_c$'s of samples A, B, and C, as determined by the midpoint of the resistivity drop, are 7 K, 10 K, and 4 K, respectively. This trend is consistent with data for similar films reported before and it reflects the systematic change of $T_c$ with the increase of $h_{Se/Te}$.\cite{huang:10}

Upon application of hydrostatic pressure the resistivity drop of samples A, B, C shifts roughly in parallel toward higher temperature, i.e. $T_c$ increases for all three samples. The results of the pressure dependent resistivity measurements are shown in Fig. 2 (a to d). While pressure does increase $T_c$ of all superconducting films, the rate of increase is very different. Although the ambient pressure $T_c$'s of samples A, B, and C differ by more than 100 \%, their $T_c$'s at the highest pressure of this investigation are relatively close at 13 K, 16 K, and 12.5 K, respectively. The largest (relative) change is observed in film C ($T_s$=340$^o$C) with an increase of $T_c$ by a factor of three. The width of the superconducting transition, as defined by the temperature difference between the 10 \% and 90 \% resistivity drop, is of the order of 1.5 K to 2.5 K, in reasonable agreement with ambient pressure data obtained for thin films\cite{tsukada:10} and high-quality single crystals.\cite{sales:09,taen:09,yeh:09}

\begin{figure}
\includegraphics[width = 3in]{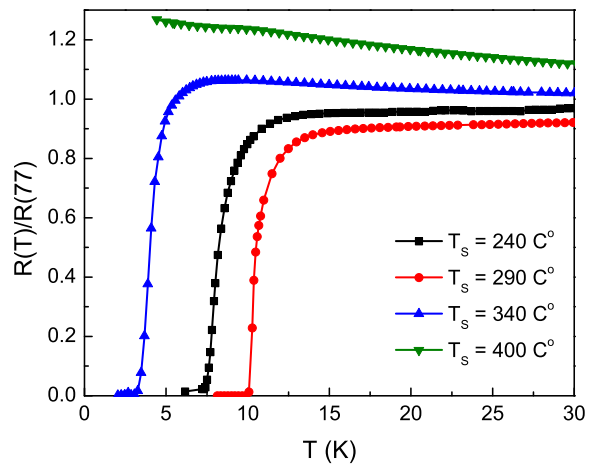}
\caption{(Color online) Resistivity vs. temperature of strained FeSe$_{0.5}$Te$_{0.5}$ thin films deposited at different substrate temperatures $T_s$.}
\end{figure}
\begin{figure}
\includegraphics[width = 3.3in]{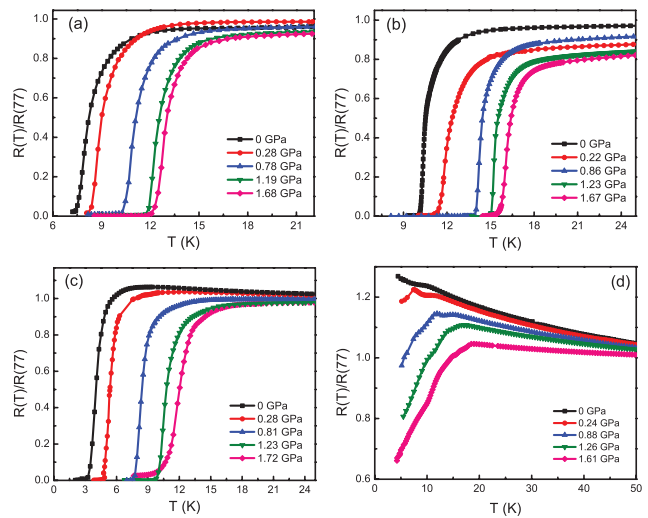}
\caption{(Color online) Pressure effect on the resistivity of the FeSe$_{0.5}$Te$_{0.5}$ films. (a) $T_s$=240 $^o$C, (b) $T_s$=290 $^o$C, (c) $T_s$=340 $^o$C, and (d) $T_s$=400 $^o$C.}
\end{figure}

In contrast, film D maintains its semiconducting temperature characteristics although the slope of the resistivity $\rho(T)$ clearly decreases with pressure (Fig. 2d). A drop of the resistivity develops at pressures above 0.8 GPa indicating the possible onset of superconductivity in this film. However, the resistance drop at 4.2 K reaches only 30 \% of the value at $T_c$ and the transition appears to be very broad at the maximum pressure of this study. This indicates that the superconducting state may be filamental and a percolating superconducting path is not achieved throughout the film. It is possible that bulk superconductivity in sample D requires much higher pressure values.

The pressure dependence of $T_c$ for the three superconducting films A, B, and C is shown in Fig. 3a. Up to the maximum pressure of this study, $T_c$ increases continuously with external pressure. The $T_c$ of film B, starting with the highest ambient value of 10.5 K, rises to 16.5 K at 1.7 GPa. This value is comparable with the $T_c$ of bulk FeSe$_{0.5}$Te$_{0.5}$ at about the same pressure,\cite{horigane:09,tsoi:09} although the ambient pressure $T_c$ of this film is clearly lower than that of the bulk compound (13 K). The critical temperatures of films A and C increase with pressure but remain lower at 1.7 GPa since the ambient pressure values are smaller. However, plotting the relative change, $T_c(p)/T_c(0)$, Figure 3b shows a significantly larger pressure effect on the normalized $T_c$ of sample C. The critical temperature increases by a factor of three upon compression of the film. The relative change of $T_c$ is approximately the same for films A and B.

The results of the high pressure study of the three superconducting films raise the question about the most relevant parameters affected by pressure. The anion height $h_{Se/Te}$ is apparently an important control parameter for $T_c$ at ambient conditions, as shown for the current films\cite{huang:10} and many other iron pnictides.\cite{mizuguchi:10} For bulk samples of FeSe, detailed structural studies at high pressure have revealed that the anion height $h_{Se}$ decreases with pressures up to 2 GPa (the pressure range of the current experiment).\cite{margadonna:09} A strict correlation of $T_c$ and $h_{Se}$ was demonstrated and discussed in high-pressure experiments on FeSe.\cite{okabe:10} The effect of hydrostatic pressure on $h_{Se/Te}$ of films attached to a substrate, however, is more complex and needs a detailed discussion.

While pressure is applied uniformly to the sample through the liquid pressure medium, the structural distortion of the FeSe$_{0.5}$Te$_{0.5}$ thin films will be affected by the MgO substrate, different from the compression of bulk samples. A qualitative picture can be derived by comparing the compressibilities of the film and substrate materials. The compressibility of MgO is $\kappa_{MgO}$=0.0154 GPa$^{-1}$.\cite{goble:85} Detailed structural data at high pressure are not available for FeSe$_{0.5}$Te$_{0.5}$ but it is justified to use compressibility data for FeSe instead. The compressibility (below 2 GPa) for FeSe is about two times larger ($\kappa_{FeSe}$=0.032 GPa$^{-1}$)\cite{braithwaite:09,garbarino:09} than that of MgO. The attachment of the FeSe$_{0.5}$Te$_{0.5}$ films to the less compressible MgO substrate results in a relatively larger uniaxial strain effect along the c-axis (perpendicular to the substrate) of the film as compared to bulk samples under the same pressure. It appears conceivable to conclude that the reduction of the anion height of the films on MgO at a given pressure is even amplified with respect to the similar effect in bulk single crystals.

It seems therefore natural to consider the compression of the tetrahedra and the decrease of $h_{Se/Te}$ with pressure as a driving mechanism to the observed raise of $T_c$. This effect is certainly essential to understand the huge increase of $T_c(p)/T_c(0)$ of film C (Fig. 3b). This film starts with the largest $h_{Se/Te}$ of the superconducting samples at ambient pressure, significantly larger than the optimal anion height. Therefore, the reduction of $h_{Se/Te}$ under pressure moves the anion closer to the optimal distance from the Fe plane and results in a large positive pressure effect on $T_c$.

\begin{figure}
\includegraphics[width = 3.3in]{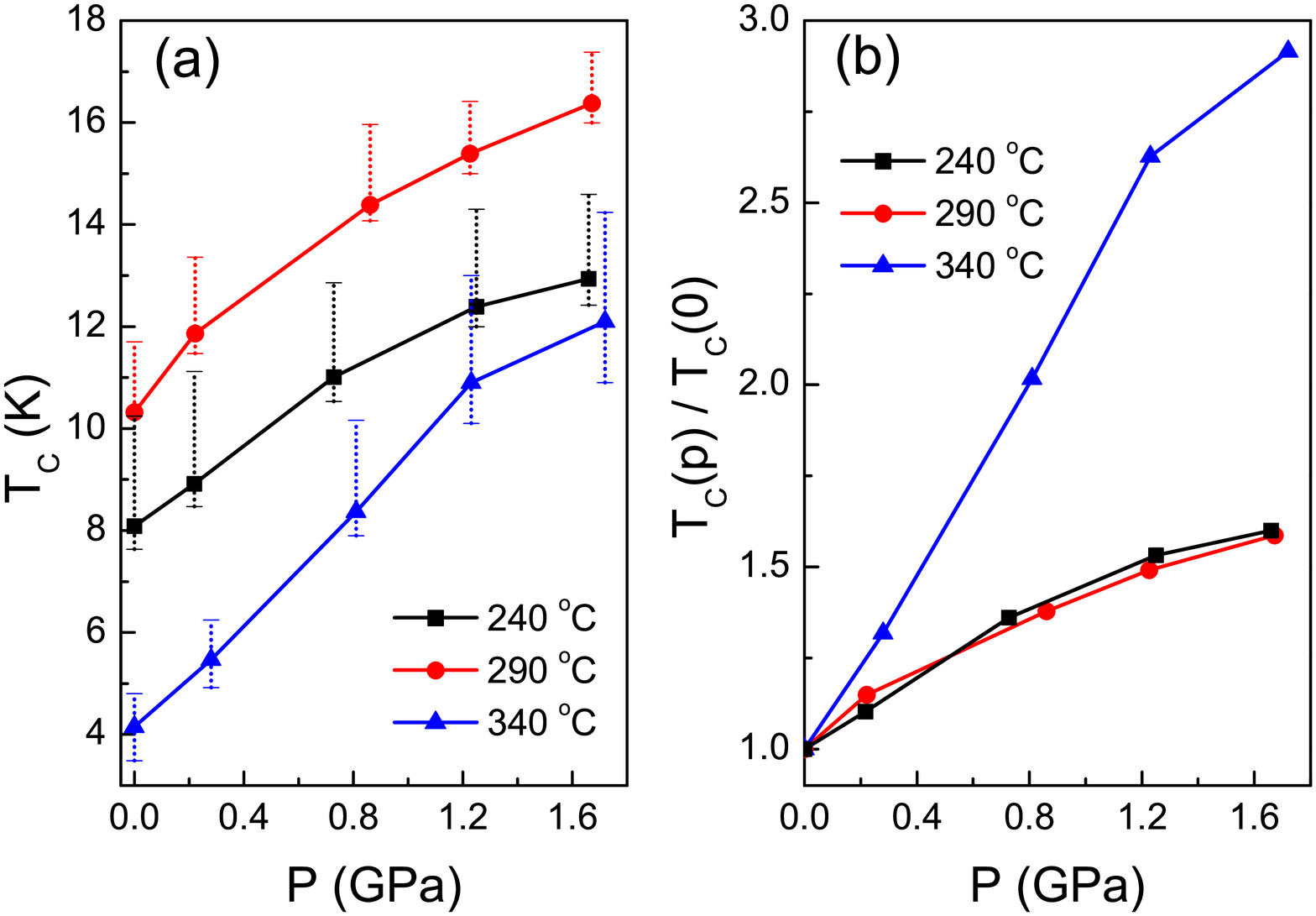}
\caption{(Color online) Pressure dependence of (a) the critical temperatures $T_c$ and (b) the relative changes $T_c(p)/T_c(0)$ for the three superconducting films. In (a) the width of the transition as defined between the 10 \% and 90 \% drop of resistivity is indicated by vertical dotted bars.}
\end{figure}

This argument, however, cannot solely account for the $T_c$ increase of sample A for which $h_{Se/Te}$ actually is already lower than the optimal value at ambient pressure. A further compression should be counterproductive in raising the superconducting $T_c$.\cite{okabe:10} Therefore, external pressure must also change other parameters favoring the superconducting state and a higher $T_c$. Changes of the band structure and the Fermi surface may play an important role in understanding the pressure effects in iron pnictides and chalcogenides. For example, the density of states, the position of the Fermi energy, the topology of the Fermi surface and the possible nesting feature can all be affected by external pressure. A charge transfer from the chalcogenide to the iron layer could also be induced by pressure. We conclude that the effect of pressure on $T_c$ of iron chalcogenides cannot be reduced to only one parameter ($h_{Se/Te}$), as suggested from high pressure studies of bulk samples,\cite{okabe:10} and further studies are needed for a better understanding.

The pressure data for the resistivity of the semiconducting film D (Fig. 2d) indicate the possible onset of superconductivity with the drop of the resistivity at higher pressures. While the temperature dependence of the resistivity is still semiconducting above this temperature, the decrease of the resistivity by 50 \% is not small. The change of the temperature dependence of the resistivity with pressure suggests that a metallic state can possibly achieved at higher values of pressure. It can be expected that this will also result in a zero resistance state and bulk superconductivity at low temperature.

\section{Summary and conclusions}
The study of the effect of pressure on prestrained thin films of FeSe$_{0.5}$Te$_{0.5}$ reveals a complex behavior of the superconducting critical temperature. While the $T_c$ of all films increases from their ambient pressure values, the relative increase of $T_c(p)/T_c(0)$ is significantly larger for the film C with the largest anion height above the Fe plane at ambient pressure. The results suggest that the effect of pressure on reducing the anion height is essential to understand the behavior of film C. However, this effect cannot solely explain the $T_c$ increase with pressure of films B and A with an initial anion height closer to or even below the optimal value. Therefore, the pressure effects on additional parameters related to the electronic structure and a possible charge transfer and the corresponding change of $T_c$ have to be considered.

%
%
%
\begin{acknowledgments}
This work is supported in part by the T.L.L. Temple Foundation, the John J. and Rebecca Moores Endowment, the State of Texas through TCSUH, the U.S. Air Force Office of Scientific Research, and the the U.S. Department of Energy.
\end{acknowledgments}

%


\end{document}